\documentclass[twocolumn,preprintnumbers, amssymb,amsmath,aps,prd,floatfix,nofootinbib,superscriptaddress,showpacs]{revtex4-1}

\usepackage{epsfig}
\usepackage{amssymb}
\usepackage{amsmath}
\usepackage{lipsum}
\usepackage{graphicx}
\usepackage[dvipsnames]{xcolor}
\usepackage{tikz}
\usepackage{bm}
\usepackage{subfigure}
\usepackage[
    colorlinks,
    linkcolor=blue,
    anchorcolor=black,
    citecolor=blue
]{hyperref}
\usepackage{soul}
\usepackage[version=4]{mhchem}
\allowdisplaybreaks[4]

\begin{document}

\title{Unified Resummation of Soft Gluon Radiation in Heavy Meson Pair Photoproduction}

\author{Cyrille Marquet} 
\email{cyrille.marquet@polytechnique.edu} 
\affiliation{CPHT, CNRS, \'Ecole polytechnique,  Institut Polytechnique de Paris, 91120 Palaiseau, France}

\author{Yu Shi} 
\email{yu.shi@polytechnique.edu}
\affiliation{CPHT, CNRS, \'Ecole polytechnique,  Institut Polytechnique de Paris, 91120 Palaiseau, France}
\affiliation{Key Laboratory of Particle Physics and Particle Irradiation (MOE), Institute of frontier and interdisciplinary science, Shandong University, Qingdao, Shandong 266237, China}

\author{Bo-Wen Xiao}  
\email{xiaobowen@cuhk.edu.cn}
\affiliation{School of Science and Engineering, The Chinese University of Hong Kong (Shenzhen), Longgang, Shenzhen, Guangdong, 518172, P.R. China}
\affiliation{Southern Center for Nuclear-Science Theory (SCNT), Institute of Modern Physics, Chinese Academy of Sciences, Huizhou, Guangdong 516000, China}

\begin{abstract} 
We develop a unified resummation framework for heavy-meson pair photoproduction that treats soft-gluon radiation in a massive scheme for $|\boldsymbol q| \lesssim m_Q$ and is consistent with the massless limit $m_Q \ll |\boldsymbol q|$, where $\boldsymbol q$ denotes the transverse momentum of the pair and $m_Q$ the quark mass. This framework describes the full correlation regime $|\boldsymbol q| \ll |\boldsymbol P|$, with $\boldsymbol P$ the pair's relative transverse momentum. Next, within the Color Glass Condensate framework, we quantify the impact of soft gluon radiation on two important phenomenological observables, namely, the azimuthal $\Delta\phi$ correlation and the harmonic asymmetries $\langle \cos(2n\phi_{qP})\rangle$. Our results show that the resummation provides an excellent description of the H1 $\Delta\phi$ data. We then find that the angular-asymmetry harmonics are sizable and their ratios in $\gamma A$ to $\gamma p$ collisions are highly sensitive to gluon saturation. We further find a clear mass hierarchy in both observables, indicating that heavy-quark masses suppress soft radiation and thereby enhance genuine small-$x$ saturation signals. Our work demonstrates that heavy-quark pair photoproduction provides a quantitative probe of gluon saturation, directly measurable in forthcoming EIC and UPC measurements. Finally, the unified resummation is readily extendable to other processes involving massive particles and jets in both $eA$ and $pA$ collisions. 
\end{abstract}

\maketitle

\textit{Introduction---} The gluon saturation phenomenon, described in QCD by means of the Color Glass Condensate (CGC) effective theory, arises from the non-linear small-$x$ evolution of extremely dense gluon fields in the high-energy limit~\cite{Gribov:1983ivg,Mueller:1985wy,Mueller:1989st,McLerran:1993ni,McLerran:1993ka,McLerran:1994vd,Iancu:2003xm,Gelis:2010nm}. Meanwhile, a precise determination of the non-perturbative (NP) gluon transverse-momentum–dependent (TMD) distributions, in particular the Weizsäcker–Williams (WW) distribution \cite{Kharzeev:2003wz,Dominguez:2010xd}, and the quantitative understanding of their non-linear small-$x$ evolution, have become central objectives of high-energy QCD~\cite{Dominguez:2011wm}. These goals are of increasing importance with the advent of upcoming high-luminosity facilities, notably the Electron–Ion Collider (EIC)~\cite{Boer:2011fh,Accardi:2012qut,Proceedings:2020eah,AbdulKhalek:2021gbh,AbdulKhalek:2022hcn}.

Using two-particle measurements, in the back-to-back correlation limit the transverse momentum imbalance of pair $\boldsymbol q$ is much smaller than the relative momentum of pair $\boldsymbol P$ and, at leading order (LO), directly tracks the intrinsic gluon transverse momentum. Thus two-particle production (e.g., dijet, dihadron, or heavy meson pairs) provides a clean channel for accessing gluon TMDs and studying the impact of saturation effects on their $x$ and $\boldsymbol q$ dependence. It is known however that
in the correlation limit the effects of soft gluon radiation are equally important, and those can potentially mask saturation signals. In the letter, we study the case of heavy-quark pair production, and we demonstrate that their large mass suppresses soft radiation and therefore allows small-$x$ saturation signals to emerge.

We shall focus on photon-initiated processes which are sensitive to the WW distribution.
On the experimental side, two-particle production involving D mesons was studied at HERA~\cite{ZEUS:2005qyx,H1:2006alo, H1:2011myz}, and the CMS Collaboration has recently announced a new program to measure charm-quark dijet/dihadron production in ultra-peripheral collisions (UPCs) with the high-statistics Run $3$ dataset~\cite{Innocenti_D0_UPC_DiffLowx2024_tr, CMS:2025jjx}. At the future EIC, precise measurements of heavy-meson pairs will enable us to probe WW gluon TMD and to search for experimental signatures of gluon saturation at small-$x$.  

Two key classes of observables for doing TMD tomography with two-particle correlations are: (i) azimuthally symmetric distributions such as the $\Delta\phi$ distribution, and (ii) azimuthal asymmetries $\langle \cos(2n\phi_{qP})\rangle$, with $\phi_{qP}$ the angle between $\boldsymbol q$ and $\boldsymbol P$. These observables also serve as prime channels for probing experimental signatures of gluon saturation, and have been widely studied in both the inclusive processes and diffractive processes, from $eA$~\cite{Mueller:2012uf,Mueller:2013wwa,Zheng:2014vka,Altinoluk:2015dpi, Metz:2011wb,Dominguez:2011br,Dumitru:2015gaa,Dumitru:2016jku,Boer:2016fqd,Dumitru:2018kuw,Hatta:2016dxp,Altinoluk:2015dpi,Kotko:2017oxg,Mantysaari:2019csc,Boussarie:2019ero,Salazar:2019ncp,Bergabo:2021woe,Zhao:2021kae,Boussarie:2021ybe,Boer:2021upt,Hagiwara:2021xkf,Iancu:2021rup,Taels:2022tza,Hatta:2022lzj,Caucal:2022ulg,Tong:2022zwp,Caucal:2023fsf,Tong:2023bus,Rodriguez-Aguilar:2023ihz,Caucal:2023nci,Shao:2024nor,Caucal:2024nsb} to $pA$ collisions~\cite{Kharzeev:2004bw,Jalilian-Marian:2005qbq,Marquet:2007vb,Stasto:2011ru,Akcakaya:2012si,Jalilian-Marian:2012wwi,Kutak:2012rf,Lappi:2012nh,Rezaeian:2012wa,Stasto:2012ru,Kovner:2014qea,vanHameren:2014ala,Kotko:2015ura,Basso:2015pba,Kovner:2015rna,Basso:2016ulb,Rezaeian:2016szi,vanHameren:2016ftb,Boer:2017xpy,Hagiwara:2017fye,Benic:2017znu,Albacete:2018ruq,Stasto:2018rci,Marquet:2019ltn,vanHameren:2019ysa,Goncalves:2020tvh,Kolbe:2020tlq,vanHameren:2020rqt,Benic:2022ixp,Al-Mashad:2022zbq}. For massive particles, soft-gluon radiation in the $ m,|\boldsymbol q|\ll |\boldsymbol P|$ limit has been investigated in Refs.~\cite{Catani:1990eg,Zhu:2013yxa, Klein:2018fmp,Klein:2020jom, delCastillo:2020omr, Ghira:2023bxr, Shi:2024gex, Ghira:2024nkk, Aglietti:2024zhg, Catani:2014qha,Catani:2017tuc,Hatta:2021jcd, Shao:2023zge}; it is known to generate non-trivial and sizable harmonics~\cite{Catani:2014qha,Catani:2017tuc,Hatta:2021jcd, Shao:2023zge}, while its impact on azimuthally symmetric distributions has also been investigated~\cite{Zhu:2013yxa, Klein:2018fmp,Klein:2020jom, delCastillo:2020omr, Ghira:2023bxr, Shi:2024gex, Ghira:2024nkk, Aglietti:2024zhg}. Interestingly enough, in the context of $\gamma\gamma\to l^+ l^-$~\cite{Klein:2018fmp,Klein:2020jom,Hatta:2021jcd,Shao:2023zge,Shi:2024gex}, the $\Delta\phi$ distribution and azimuthal asymmetric are obtained using distinct Sudakov factors, hereby posing the challenge of a unified resummation description. In this Letter, we combine the massive scheme for $|\boldsymbol q| \lesssim m$ and the massless scheme for $m \ll |\boldsymbol q|$, with a smooth crossover, and obtain a unified resummation framework that consistently describes both types of observables. This is one of our main achievements.

The key physics consideration for the unified resummation framework centers on the interplay among three characteristic scales in heavy-quark pair production: $\boldsymbol q$, $\boldsymbol P$, and the quark mass $m_Q$. It must account for the scale hierarchy between $\boldsymbol q$ and $m_Q$, which leads to two qualitatively different regimes. (i) When $|\boldsymbol q| \lesssim m_Q$, the heavy-quark mass acts as a natural infrared cutoff, regulating collinear singularities and implementing the dead-cone suppression of soft gluon emission, yielding the logarithm involving the mass effect $\alpha_s/\boldsymbol{q}^2\ln(\boldsymbol{P}^2/m_Q^2)$. (ii) In the small-mass regime with $m_Q \ll |\boldsymbol q|$, the resummation is insensitive to the mass scale and behaves like a light-quark system, with the corresponding logarithm  $\alpha_s/\boldsymbol{q}^2\ln(\boldsymbol{P}^2/\boldsymbol{q}^2)$.

Our new resummation framework provides the first quantitative description of the H1 data for the $\Delta\phi$ distribution. Using the CGC framework, we further carry out an analysis of azimuthal $\Delta \phi$ correlations and asymmetries in UPCs and at the EIC.
We find that these observables are highly sensitive to saturation dynamics, establishing heavy-meson pair photoproduction as a powerful new probe of gluon saturation for the future EIC and UPC programs.

\textit{Soft gluon radiation and unified resummation---} Let us start with the inclusive heavy-quark pair photoproduction process, $ \gamma + g  \rightarrow  Q + \bar{Q}+ X$ where the heavy quarks then hadronize into a heavy-meson pair. At one-loop order, the process $\gamma(l)+g(p)\rightarrow Q(k_{1})+\bar{Q}(k_{2})+\text{\ensuremath{g(k_{g})}}$ yields a cross section that can be written as
\begin{eqnarray}
\frac{d\sigma }{d^2\boldsymbol{P} d^{2}\boldsymbol{q}} \! =\! \int d^2\boldsymbol{q}' d^2\boldsymbol{k}_g \frac{d\sigma_{\rm LO}}{d^2\boldsymbol{P}d^{2}\boldsymbol{q}' } \mathcal{S}_{r}(\boldsymbol{k}_g)\delta^{(2)}\left(\boldsymbol{q}-\boldsymbol{k}_g -\boldsymbol{q}'  \right),
\end{eqnarray}
where $\sigma_{\rm LO}$ stands for the LO cross section. In the back-to-back region ($|\boldsymbol{q}| \ll |\boldsymbol{P}| \simeq |\boldsymbol{k}_1|  \simeq|\boldsymbol{k}_2|$ with $\boldsymbol{q}=\boldsymbol{k}_1\!+\!\boldsymbol{k}_2 $), the eikonal radiation kernel $\mathcal{S}_{r}(\boldsymbol{k}_g)$ is given by
\begin{align}
&\mathcal{S}_{r}(\boldsymbol{k}_g) = \frac{\alpha_{s}}{2\pi^{2}}\int dy_g \Big[\frac{N_c}{2}
 \frac{2k_{1}\cdot p}{(k_{1}\cdot k_{g})(p\cdot k_{g})}\nonumber \\
&+\frac{N_c}{2}\frac{2k_{2}\cdot p}{(k_{2}\cdot k_{g})(p\cdot k_{g})} -\frac{1}{2N_{c}} \frac{2k_{1}\cdot k_{2}}{(k_{1}\cdot k_{g})(k_{2}\cdot k_{g})}\Big]. 
\end{align} 
Following techniques developed in Ref.~\cite{Hatta:2021jcd}, it can be decomposed as follows 
\begin{equation}
\mathcal{S}_{r}(\boldsymbol{k}_g) = \mathcal{S}_{r0}(\boldsymbol{k}_g)+2\sum_{n=1}^{\infty}\cos(2n\phi_{k_gP})\mathcal{S}_{r2n}(\boldsymbol{k}_g), 
\end{equation}
where the Fourier coefficients read
\begin{equation}
\mathcal{S}_{r2n}(\boldsymbol{k}_g) = \int^{2\pi}_0 d\phi_{k_gP} \cos(2n\phi_{k_gP}) \mathcal{S}_{r}(\boldsymbol{k}_g). 
\label{eq::erf}
\end{equation}

First, let us carry out the calculation of the massive scheme in the regime  $|\boldsymbol{k}_g|\leq m_Q$. The mass of the particles serves as a natural regulator, eliminating collinear divergences from the above expression. Assuming $m_Q\ll |\boldsymbol{P}|$, we can express the soft radiation contribution as
\begin{align}
&\mathcal{S}_{r}(\boldsymbol{k}_g)    = \frac{\alpha_{s}}{2\pi^{2}}\frac{1}{\boldsymbol{k}_g^{2}}\Bigg[N_{c}\ln\frac{M_{\rm in}^{2}}{\boldsymbol{k}_g^{2}}-N_{c}\ln\frac{M_{\rm in}^{2}}{\boldsymbol{P}^{2}}\nonumber \\
&+2C_{F}\left(c_{0}(m_Q)+2\sum_{n=1}c_{2n}(m_Q) \cos\left(2n\phi_{k_gP}\right)\right)\Bigg],
\label{eq::s1}
\end{align}
where $M_{\rm in}^{2}\equiv (k_1+k_2)^2 \simeq 2\boldsymbol{P}^2(1+\cosh \Delta y)$ stands for the invariant mass of the heavy quark pair with the rapidity difference $\Delta y=|y_1-y_2|$. The invariance under quark–antiquark interchange eliminates all odd harmonic coefficients $c_{2n-1}$. Thus, the non-zero coefficients $c_{2n}$ (for $n=0,1,2,3$) can be expressed as
\begin{align}
c_{0} (m_Q) = \ln\frac{M_{\rm in}^{2}}{m_{Q}^{2}}, & \quad  c_{2n} (m_Q) = \ln\frac{M_{\rm in}^{2}}{m_{Q}^{2}}+f_{2n}(\Delta y).
\end{align}
 The Detailed expression for $f_{2n}(\Delta y)$ is provided in the Supplemental Material (SM). Our results are consistent with Eq. (50) of Ref.~\cite{Hatta:2021jcd} when we make the substitution $m_Q^{2}\!\to\! \boldsymbol{ P}^{2}R^{2}$ in the coefficients $c_{0}$ and $c_{2}$, where the effective “jet mass’’ ($\sim\boldsymbol{P}R$) acts as a kinematic collinear regulator. In the massless limit $m_Q \rightarrow 0$, applying a jet algorithm produces the same logarithmic structure, with $m_Q^{2}$ simply replaced by $\boldsymbol P^{2}R^{2}$. 

Second, in the regime where $m_Q \ll |\boldsymbol{k}_g|$, heavy quarks effectively behave as massless particles; therefore, we take the $m_Q \to 0$ limit in the following calculation. To illustrate how the large logarithms emerge, we examine the first term in the integral over the eikonal radiation functions given in Eq.~(\ref{eq::erf}):
\begin{eqnarray}
 \frac{ \alpha_{s} N_c}{2\pi^2} \int d\phi_{k_gP} \int\frac{d\xi}{\xi}\frac{z^2}{\left(z\boldsymbol{k}_{g}-\xi\boldsymbol P\right)^{2}},
\end{eqnarray}
where $z=k^+_1/l^+$ and $\xi=k^+_g/l^+$ denote the longitudinal plus momentum fractions of the heavy quark and the emitted gluon with respect to the incoming photon, with $l^+$ being the photon’s plus momentum. The denominator is related to the three-dimensional angular separation between the heavy quark and the emitted gluon as follows
\begin{eqnarray}
\left(\xi\boldsymbol{P}-z\boldsymbol{k}_{g}\right)^{2}  = \xi z\boldsymbol P \boldsymbol{k_{g}}R_{1g}^{2}<z^2\boldsymbol P^{2}R_{1g}^{2},
\end{eqnarray}
with $R_{1g}^{2}=2 \cosh{\Delta y_{1g}}-2\cos \phi_{k_gP}\approx\Delta y_{1g}^{2}+\Delta\phi_{k_gP}^{2}$ for small separations. Collinear divergences arise when $z\boldsymbol{k}_{g}=\xi\boldsymbol{P}$, and they are normally absorbed into the renormalized fragmentation function, yielding a finite short-distance coefficient. For the Sudakov large logarithms, we remove the collinear divergence by taking the $\xi \to 0$ limit in the denominator~\cite{Mueller:2013wwa}, which yields $\left(\xi\boldsymbol{P}-z\boldsymbol{k}_{g}\right)^{2}\approx z^{2}\boldsymbol{k}_{g}^{2}$.
 Combining these kinematic relations, one obtains the constraint $\boldsymbol{k}_g^2 < \boldsymbol P^{2}R_{1g}^{2}$ for isolating large logarithms. After implementing this kinematic constraint in Eq.~(\ref{eq::erf}), one obtains
\begin{align}
&\mathcal{S}_{r}(\boldsymbol{k}_g)    = \frac{\alpha_{s}}{2\pi^{2}}\frac{1}{\boldsymbol{k}_g^{2}}\Bigg[N_{c}\ln\frac{M_{\rm in}^{2}}{\boldsymbol{k}_g^{2}}-N_{c}\ln\frac{M_{\rm in}^{2}}{\boldsymbol{P}^{2}}\nonumber \\
&+2C_{F}\left(c_{0}(\boldsymbol{k}_g)+2\sum_{n=1} c_{2n}(\boldsymbol{k}_g) \cos\left(2n\phi_{k_gP}\right)\right)\Bigg],
\label{eq::s2}
\end{align}
which is identical to Eq.~(\ref{eq::s1}) except that the coefficients $c_{0}$ and $c_{2n}$ are evaluated at $\boldsymbol{k}_g^{2}$ instead of $m_Q^2$.

In the following, we unify these two schemes to cover the full $m_Q,|\boldsymbol{q} |\!\ll \! |\boldsymbol{P}|$ regime. After performing the all-order soft-gluon resummation in the coordinate space ($\boldsymbol{k}_g \to \boldsymbol{b}$), we arrive at the resummed cross section for the hadron-level pair production
\begin{align}
&\frac{d\sigma}{d\Omega} =
 \int \frac{dz_{1h}}{z_{1h}^2} \frac{dz_{2h}}{z_{2h}^2} \int\frac{d^{2}\boldsymbol{b} }{(2\pi)^{2}}e^{i\boldsymbol{q}\cdot\boldsymbol{b}}
\mathcal{H}_{ij}(z,\boldsymbol{P}) \widetilde  W^{ij}(x_g, \boldsymbol{b})
\nonumber \\
  \times  &\Big[1+\sum_{n=1} (-1)^n \frac{2C_{F}\alpha_{s} (\mu_b) C_{2n}(\mu_{b}) }{n\pi}\cos(2n\phi_{bP})\Big],
 \label{eq::res}
\end{align}
where $\mu_{b}\equiv2e^{-\gamma_{E}}/|\boldsymbol{b}|$ and $d\Omega\!=\!dy_{1}dy_{2}d^{2}\boldsymbol{P}_h d^{2}\boldsymbol{q}_h$. $\boldsymbol{p}_{i}$ and $y_i$ denote the transverse momentum and rapidity of each heavy meson, respectively, and we defined $\boldsymbol{P}_h\!=\!(\boldsymbol{p}_1\!-\!\boldsymbol{p}_2)/2$ and $\boldsymbol{q}_h=\boldsymbol{p}_1+\boldsymbol{p}_2$. The meson momenta are related to the parent heavy-quark momenta $\boldsymbol{k}_i$ via $\boldsymbol{p}_{i}=z_{ih}\boldsymbol{k}_{i}$, where $z_{ih}$ denotes the meson's momentum fraction. $\mathcal{H}_{ij}$ stands for the LO TMD hard factor and the $\widetilde W$ function is defined as
\begin{equation}
\widetilde W^{ij} = xG^{ij}(x_g, \boldsymbol{b})  D_{h_1/Q}(z_{1h}, \mu_b)D_{h_2/\bar Q}(z_{2h}, \mu_b) e^{-{\rm Sud}_{}(\boldsymbol{b})}, 
\end{equation}
where $ D_{h/Q}$ represents the heavy-quark fragmentation functions. $xG^{ij}$ is the WW TMD gluon distribution, which includes both the unpolarized and linearly polarized parts.
The Sudakov factors include both perturbative and non-perturbative parts: 
${\rm Sud}_{\rm}=N_c\,{\rm Sud}_{i}+2C_F\,{\rm Sud}^{Q}_{f}+{\rm Sud}_{\rm NP}$.

The perturbative part due to the initial gluon radiation results from the resummation of the "$N_c$" term of $ \int d^2\boldsymbol{k}_g\ \mathcal{S}_{r}(\boldsymbol{k}_g)[e^{i \boldsymbol{k}_g\cdot\boldsymbol{b}}-\theta(M_{\rm in}^2-\boldsymbol{k}_g^2)]$ (with the virtual contribution included) and is given by
\begin{equation}
{\rm Sud}_{i} = \int_{\mu_{b}^{2}}^{M_{{\rm in}}^{2}}\frac{d\mu^{2}}{\mu^{2}}\frac{\alpha_{s}(\mu)}{2\pi}
\left[
\ln\frac{ M^2_{\rm in} }{\mu^{2}}-
\ln\frac{ M^2_{\rm in}}{\boldsymbol{P}^{2}}
\right].
\label{eq::ISF}
\end{equation}
The initial double Sudakov factor is consistent with the massless case~\cite{Hatta:2021jcd,Shao:2024nor}. Similarly, the perturbative part computed from the final-state radiation off the heavy-quark pair originates from the resummation of the "$c_0$" terms, unified as $c_0(m_Q)\theta(m_Q\!-\!\boldsymbol{k}_g) + c_0(\boldsymbol{k}_g)\theta(\boldsymbol{k}_g\!-\!m_Q)$ inside the $\boldsymbol{k}_g$ integral, and is given by
\begin{eqnarray}
{\rm Sud}^{Q}_{f}\! =\! \int_{\mu_{b}^{2}}^{M_{{\rm in}}^{2}}\!\frac{d\mu^{2}}{\mu^{2}}\frac{\alpha_{s}(\mu)}{2\pi} \Bigg[ \ln\frac{ M_{\rm in}^{2} }{\mu^{2}} -\theta(m_Q^2\!-\!\mu_b^2)\ln\frac{ m_Q^2 }{\mu^{2}} \Bigg].
\end{eqnarray}
After replacing the running coupling and removing the color factor $C_F$, the final-state Sudakov factor becomes identical to the QED lepton Sudakov factor, as obtained in the azimuthally symmetric case~\cite{Klein:2020jom,Shi:2024gex}. It shows that this unified resummation scheme consistently captures both azimuthally symmetric and asymmetric observables. When $m_Q \gtrsim \mu_b$, the quark mass becomes essential, as it removes collinear final-state radiation (the dead-cone effect) and renders the Sudakov factor single-logarithmic, similar to the QED lepton case observed in azimuthally asymmetric distributions~\cite{Hatta:2021jcd}. In the $m_Q \to 0$ limit, the mass effects disappear and the Sudakov factor reduces to the familiar double-logarithmic form, identical to the first-term Sudakov factor from initial-state gluon radiation in Eq.~(\ref{eq::ISF}).

Note that, unlike the rest of the expression, the "$c_{2n}$" part of $\mathcal{S}_{r}$ (see Eq.~\eqref{eq::s1} and Eq.~\eqref{eq::s2}) is not accompanied by virtual terms. Nevertheless, the result of the Fourier transformation is still finite due to the additional $\cos(2n\phi_{kgP})$ \cite{Catani:2017tuc,Hatta:2021jcd}. Moreover, it does not contain large logarithms and need not be resummed. In coordinate space, their unification can be written in terms of the following coefficients:
\begin{eqnarray}
C_{2n}(\mu_{b})=\ln\frac{M_{{\rm in}}^{2}}{\mu_{b,2n}^{2}}-\ln\frac{m_Q^{2}}{\mu_{b,2n}^{2}}\theta(m_{Q}^{2}-\mu_{b,2n}^{2})+f_{2n}(\Delta y),
\label{eq::c2n}
\end{eqnarray}
with a $n$-dependent matching point given by
$\ln\mu_{b,2n}^{2}=\ln\mu_{b}^{2}+2\psi^{(0)}(n)+2\gamma_E+1/n$, whose origin can be traced back to the momentum space matching at $m_{Q}^{2}=\boldsymbol{k}_{g}^{2}$. This approximate expression differs only marginally from the exact results presented in the SM. The first term represents the massless contribution. In the $ \mu_{b,2n}^{2} \lesssim m_Q^2$ region, the sum of the first two terms reduce to $\ln(M_{{\rm in}}^{2}/m_{Q}^{2})$.

As shown in the Supplemental Material, the unified and conventional massive-only schemes agree in the $|\boldsymbol q| < m_Q$ region but notably differ for $m_Q \ll |\boldsymbol q|$, with the unified resummation yielding sizable corrections. In addition, the deviation between the two schemes sets in earlier (at smaller ($|\boldsymbol q|$)) for higher harmonics ($n$).

Finally, to account for NP effects, we use the so-called $b_{*}$ prescription: we introduce the NP function ${\rm{Sud}}_{\rm NP}=\mathcal {C}_{\rm NP}\left[ 0.106\, \boldsymbol{b}^{2}+0.42\ln({M_{{\rm in}}}/{Q_{0}})\ln ({\boldsymbol{b}}/{b_{*}}) \right]$
where $b_{*}=|\boldsymbol{b}|/\sqrt{1+\boldsymbol{b}^{2}/b_{{\rm max}}^{2}}$ and further replace $|\boldsymbol{b}|$ by $b_{*}$ in $\mu_{b}$. We have left $\mathcal {C}_{\rm NP}$ as a free parameter while $b_{{\rm max}}=1.5$ GeV$^{-1}$ and $Q_{0}^{2}=2.4$ GeV$^{2}$~\cite{Echevarria:2020hpy,Sun:2014dqm}.

\textit{Heavy-meson pair correlations as a probe of the gluon saturation effect---}
In the CGC framework, when $|\boldsymbol{P}|\gg |\boldsymbol{q}|$, heavy-meson pair correlations are sensitive to the WW TMD gluon distribution at small-$x$, which encodes the saturation dynamics and is defined by~\cite{Dominguez:2011wm, Dominguez:2011br,Marquet:2016cgx, Marquet:2017xwy}
\begin{eqnarray}
xG^{ij}(x_{g},\boldsymbol{q}) &=&-\frac{4}{g_{s}^{2}}\int\frac{d^{2}\boldsymbol{x}d^{2}\boldsymbol{y}}{(2\pi)^{3}}e^{-i\boldsymbol{q}\cdot(\boldsymbol{x}-\boldsymbol{y})} \nonumber \\
&& \times \langle{\rm Tr}\left[(\partial^{i}U_{x})U_{y}^{\dagger}(\partial^{j}U_{y})U_{x}^{\dagger}\right]\rangle,
\label{eq:ww}\end{eqnarray}
and it can be decomposed into an unpolarized part ${xf}$ and a linearly polarized part ${xh}$~\cite{Metz:2011wb,Dominguez:2011br}
\begin{equation}
xG^{ij}(x_{g},\boldsymbol{q})
 = (\delta^{ij}/2) \, {xf}(x_{g},\boldsymbol{q}) + \Pi^{ij}(\boldsymbol{q}){xh}(x_{g},\boldsymbol{q}), 
\end{equation}
where the traceless rank-2 projector $\Pi^{ij}(\boldsymbol{q})= \boldsymbol{q}^i \boldsymbol{q}^j/ \boldsymbol{q}^2 -\delta^{ij}/2$ and $2\hat {\boldsymbol{P}}_i \hat {\boldsymbol{P}}_j \Pi^{ij}(\boldsymbol{q}) = \cos 2\phi_{qP}$. This process allows us to measure the azimuthal asymmetry $\langle \cos(2\phi_{{qP}}) \rangle$, which has long been a primary channel for directly probing the linearly polarized WW gluon distribution~\cite{Metz:2011wb,Dominguez:2011br,Dumitru:2015gaa,Marquet:2017xwy}. Let us comment on $\langle \cos(2\phi_{qP}) \rangle$ in the transverse photon channel, as it is the dominant one compared to the longitudinal case. For heavy-quark pairs produced at the same rapidity ($z=1/2$), $\langle \cos(2\phi_{qP}) \rangle$ scales as $m_Q^{2}-Q^{2}/4$ (with $Q^{2}$ the photon virtuality), and thus it is positive in low $Q^2$ events while it turns negative for $Q^{2}>4m_Q^{2}$.

For the numerical analysis, we use the JIMWLK equation~\cite{JalilianMarian:1997jx,JalilianMarian:1997gr,Iancu:2000hn,Ferreiro:2001qy} in the Gaussian approximation. Then, the WW gluon distribution Eq.~\eqref{eq:ww} can be written in terms of the adjoint dipole $S$-matrix $S_g(x_g,\boldsymbol r)$, whose square root $S$ obeys the non-linear Balitsky-Kovchegov (BK) equation \cite{Balitsky:1995ub,Kovchegov:1999yj}, see e.g.~\cite{vanHameren:2016ftb}. We solve the running-coupling BK (rcBK) evolution equation~\cite{Golec-Biernat:2001dqn,Balitsky:2006wa, Kovchegov:2006vj, Gardi:2006rp, Balitsky:2007feb,Albacete:2007yr, Albacete:2010sy,Berger:2010sh}
with the initial condition~\cite{ Albacete:2010sy} 
$S(x_{0}=0.01,\boldsymbol{r})=\exp\left[-\frac{\left(Q_{s0}^2 \boldsymbol{r}^2 \right)^{1.118} }{4}\ln\left(\frac{1}{\boldsymbol{r}\Lambda}+e\right)\right]$
using  $Q_{s0}^2=0.16$ GeV$^2$ for the proton and $Q_{s0A}^2=5\,Q_{s0}^2$~\cite{Shi:2021hwx} for the heavy nucleus. Heavy-quark hadronization is modeled using Peterson model~\cite{Peterson:1982ak}, with $m_c=1.2$ GeV and $m_b=4.5$ GeV, and we also include the effect of running coupling at one loop. Finally, all uncertainty bands are obtained by varying the NP coefficient $\mathcal {C}_{\rm NP}$ in the range $1\le \mathcal {C}_{\rm NP}\le2$.

\begin{figure}
\includegraphics[width=0.8\linewidth]{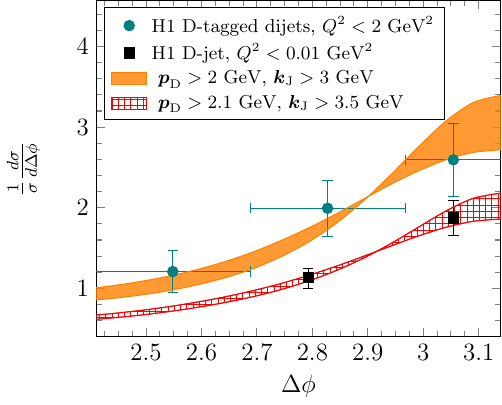}
\caption{Comparison of our resummed predictions with H1 data on the $\Delta \phi$ distribution in D-meson–triggered dijet and D-meson+jet  events~\cite{H1:2006alo,H1:2011myz}.}
\label{fig:H1_dphi}
\end{figure}

In the following, we present the numerical results. Firstly, we study the azimuthal $\Delta \phi$ correlation where $\Delta \phi=\phi_{\boldsymbol p_1}-\phi_{\boldsymbol{p}_2}$ is the angular difference between two final-state particles. The $\Delta\phi$ distributions in $D$-meson$+$jet events~\cite{H1:2006alo} and $D$-meson-triggered dijet events~\cite{H1:2011myz} were measured at HERA with a cone size $R=1$. Both measurements are treated as $D$-meson-jet correlations, for which the perturbative Sudakov factor reads
${\rm Sud}^{\rm DJet}_{\rm per}=N_c\,{\rm Sud}_{i}+C_F\,{\rm Sud}^{\rm Q}_{f}+C_F\,{\rm Sud}^{\rm Jet}_{f}$ where we obtain the jet Sudakov factor in a unified scheme following the same procedure as for the heavy-quark case. The eikonal radiation kernel  $\mathcal{S}^{\rm Jet}_{r}(\boldsymbol{k}_g)$ turns out to be identical to Eq.~\eqref{eq::s1} and Eq.~\eqref{eq::s2} with $m_Q^2$ substituted by $\boldsymbol P^{2}R^{2}$, giving:
\begin{equation}
 {\rm Sud}^{\rm Jet}_{f} \!= \!\int_{\mu_{b}^{2}}^{M_{{\rm in}}^{2}}\frac{d\mu^{2}}{\mu^{2}}\!\frac{\alpha_{s}(\mu)}{2\pi} \!\Big[ \ln\frac{ M_{\rm in}^{2} }{\mu^{2}} \!-\!\theta(\boldsymbol{P}^2\!R^2\! -\!\mu_b^2)\!\ln \!\frac{ \boldsymbol{P}^2R^2 }{\mu^{2}} \Big]. 
\end{equation}
As shown in Fig.~\ref{fig:H1_dphi}, we compare our numerical calculations with two H1 datasets on the $\Delta \phi$ distribution in the D-meson triggered dijet events~\cite{H1:2006alo, H1:2011myz}. Our results show excellent agreement with the H1 data. 

\begin{figure}
\includegraphics[width=0.8\linewidth]{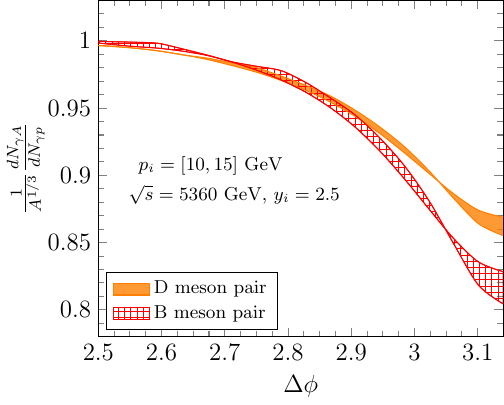}
\caption{Predictions for the nuclear modification factor
for D- and B-meson pair photoproduction in UPCs at the LHC with $\sqrt{s}=5.36$ TeV, shown as a function of $\Delta \phi$.}
\label{fig:LHC_dphi_R}
\end{figure} 
To probe gluon saturation in this channel, we compute the nuclear modification factor in $\gamma A$ relative to $\gamma p$ collisions, defined as $\mathcal{R}\equiv A^{-1/3}dN_{\gamma A}/dN_{\gamma p}$, for D- and B-meson pair photoproduction at $\sqrt{s}=5.36$ TeV (see Fig.~\ref{fig:LHC_dphi_R}). Our results show that $\mathcal{R}$ exhibits saturation-induced suppression in the deep back-to-back region ($\Delta \phi \to \pi$), with a clear mass hierarchy, $\mathcal{R}\big|_{m_c}\simeq 0.85 > \mathcal{R}\big|_{m_b}\simeq 0.80$. This behavior reflects the suppression of final-state soft radiation due to heavy-quark mass (a weaker Sudakov effect for heavier quarks), thereby enhancing the prominence of the genuine small-$x$ effects.

Furthermore, we numerically calculate the azimuthal harmonics $\langle \cos(2n\phi_{qP})\rangle$, defined by
\begin{equation}
\langle \cos (2n\phi_{qP }) \rangle = \frac{\int d \mathcal{P.S.}  \cos(2n\phi_{qP}) \frac{d\sigma}{dy_{1}dy_{2}d^{2}\boldsymbol{P}d^{2}\boldsymbol{q}} }{\int d \mathcal{P.S.}  \frac{d\sigma}{dy_{1}dy_{2}d^{2}\boldsymbol{P}d^{2}\boldsymbol{q}} }.
\end{equation}
where $d \mathcal{P.S.}$ denotes the appropriate phase‐space measure. For saturation observables, we then define the harmonic ratios $\langle \cos(2n\phi _{qP})\rangle_{\gamma A}/\langle \cos(2n\phi _{qP} )\rangle_{\gamma p}$ by comparing azimuthal asymmetries in $\gamma A$ and $\gamma p$ collisions.

\begin{figure}
\includegraphics[width=0.75\linewidth]{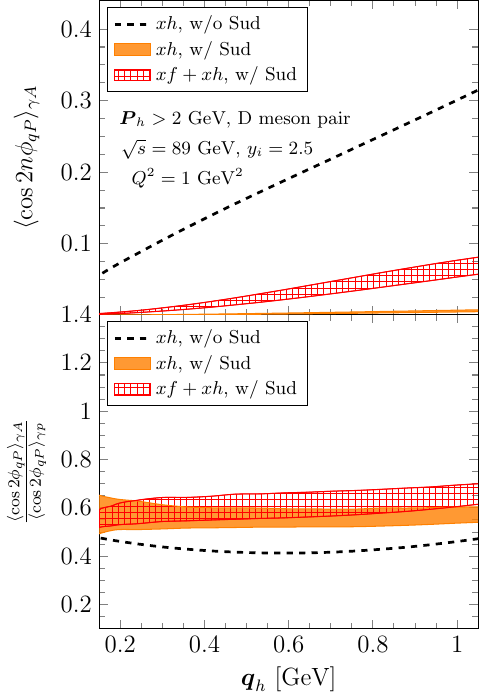}
\caption{Predictions for the azimuthal asymmetry $\langle \cos(2\phi_{qP}) \rangle$ (upper panel) and its ratio (lower panel) for D-meson pair production in the EIC regime with $Q^2=1$ GeV$^2$, plotted as a function of $\boldsymbol{q}_h$.}
\label{fig:D_cos2phi_EIC}
\end{figure} 

To access the positive, mass-dominated regime of the linearly polarized gluon contribution and to quantify soft-gluon effects, we choose EIC kinematics with a relatively small virtuality of $Q^2=1$ GeV$^2$ and present our predictions for the azimuthal asymmetry $\langle \cos2\phi_{qP} \rangle$ and the corresponding ratio for $D\bar{D}$ pair photoproduction in Fig.~\ref{fig:D_cos2phi_EIC}. As shown in the upper panel, when soft-gluon radiation is neglected, the asymmetry $\langle\cos2\phi_{qP}\rangle$ originates entirely from the linearly polarized gluon distribution $x h$. However, once multiple soft emissions are included, Sudakov effects strongly suppress the contribution from the polarized WW gluon distribution, a phenomenon also observed in $\gamma$-jet correlations in $pA$ collisions~\cite{Boer:2017xpy} and in dijet production from DIS~\cite{Zhao:2021kae,Caucal:2023fsf}. In this regime, the dominant source of $\langle\cos2\phi_{qP}\rangle$ shifts to the coupling of soft-gluon radiation with the unpolarized WW gluon distribution $x f$, which then generates a significant contribution.

As shown in the bottom panel of Fig.~\ref{fig:D_cos2phi_EIC}, although multiple-gluon radiation tends to reduce the visible impact of saturation, a strong suppression remains, with the ratio staying in the range $0.5$--$0.6$. This demonstrates that, even after including the Sudakov effects, the resulting asymmetry $\langle\cos2\phi_{qP}\rangle$ still encodes the information about the nonlinear small-$x$ dynamics and thus provides a robust and direct probe of gluon saturation.

\begin{figure}
\includegraphics[width=.75\linewidth]{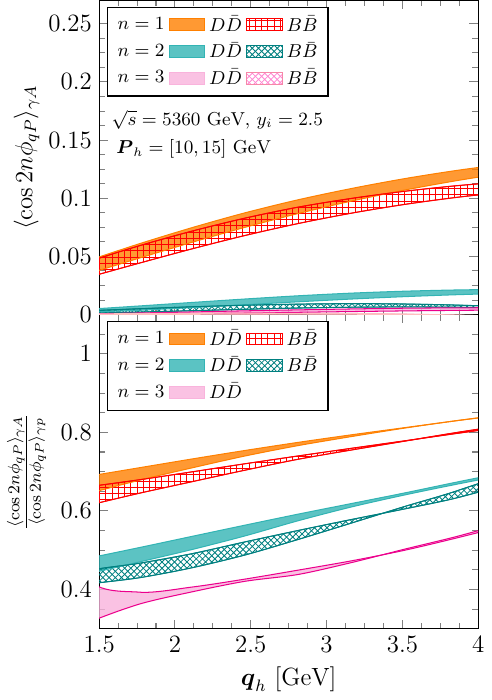}
\caption{Predicted azimuthal asymmetries $\langle \cos(2n\phi_{qP}) \rangle$ (upper) and their ratios (lower) for B- and D-meson pair photoproduction in UPCs at the LHC ($\sqrt{s}=5.36$ TeV), plotted as functions of $\boldsymbol{q}_h$.}
\label{fig:D_cos2nphi_UPC}
\end{figure}

In Fig.~\ref{fig:D_cos2nphi_UPC}, we compare the asymmetries $\langle \cos2n\phi_{qP} \rangle$ and the corresponding ratios as functions of $\boldsymbol{q}_h$ for D- and B-meson pair correlations at $\sqrt{s}=5.36$ TeV. The top panel shows that the asymmetries are sizable and exhibit a strong $n$-ordering, $\langle \cos2\phi_{qP}\rangle > \langle \cos4\phi_{qP}\rangle > \langle \cos6\phi_{qP}\rangle$. The expected mass hierarchy is barely visible due to the NP uncertainties. 

The bottom panel shows that the ratios exhibit increasingly strong suppression at higher harmonic order, $\frac{\langle \cos6\phi _{qP}\rangle_{\gamma A}}{\langle \cos6\phi _{qP} \rangle_{\gamma p}}<\frac{\langle \cos4\phi _{qP}\rangle_{\gamma A}}{\langle \cos4\phi _{qP} \rangle_{\gamma p}}<\frac{\langle \cos2\phi _{qP}\rangle_{\gamma A}}{\langle \cos2\phi _{qP} \rangle_{\gamma p}}<1$, indicating that these ratios serve as powerful observables for gluon-saturation studies. Moreover, the mass ordering resurfaces.

As detailed in the analytical discussion in the Supplemental Material, NP broadening and the perturbative Sudakov factor tend to dilute saturation signals in the lower harmonics, whereas the higher-order angular coefficients remain predominantly governed by nuclear suppression from gluon saturation. Consequently, measurements of large-$n$ azimuthal harmonics offer a unique and robust window into genuine nonlinear gluon dynamics in future EIC and UPC experiments.

\textit{Conclusion and outlook---} For heavy-meson pair photoproduction via photon–gluon fusion, we develop a soft-gluon resummation formalism that smoothly interpolates the massive and massless limits, ensuring reliable predictions across the entire $|\boldsymbol q| \ll |\boldsymbol P|$ region. This new unified framework consistently describes both azimuthally symmetric and asymmetric observables.

Within the CGC framework, our results show that our resummation scheme can accurately describe the H1 $\Delta \phi$ data for the first time.  We further demonstrated that Sudakov effects strongly suppress the linearly polarized WW gluon $xh$, whereas the unpolarized component $xf$, when coupled to multiple soft-gluon emissions, generates sizable  $\langle \cos (2n\phi_{qP}) \rangle$ harmonics. Meanwhile, the ratios of asymmetries in $\gamma A$ versus $\gamma p$ collisions decrease with harmonic order, reflecting increasingly stronger saturation-induced suppression at higher $n$. Moreover, those ratios, the $\Delta \phi$ distribution, and ${\cal R}$ exhibit a clear mass hierarchy. It reflects the mass screening of final-state soft radiation, which reduces the Sudakov effect and thereby enhances the visibility of genuine small-$x$ saturation effects. Our findings demonstrate that heavy‐quark pair photoproduction at the forthcoming EIC and in UPCs will provide a powerful new probe of gluon saturation.

Finally, the unified resummation scheme developed here is broadly applicable to processes involving massive particles and jets, opening new avenues for precision studies of soft radiation in QCD, as well as QED~\cite{Klein:2018fmp,Klein:2020jom,Hatta:2021jcd,Shao:2023zge,Shi:2024gex}.

\textit{Acknowledgements---} We thank Cheng Zhang for collaboration at the early stage of this project, and Feng Yuan, Jian Zhou, Anna M.~Sta\'{s}to, and Xuan-bo Tong for useful inputs and discussions. This work is supported in part by the Ministry of Science and Technology of China under Grant No. 2024YFA1611004, by the CUHK Shenzhen University Development Fund under Grant No. UDF01001859.

\bibliographystyle{apsrev4-1}
\bibliography{main.bib}

\newpage
\pagebreak
\begin{widetext}

\let\oldaddcontentsline\addcontentsline
\renewcommand{\addcontentsline}[3]{}
\section*{Supplemental material}

\let\addcontentsline\oldaddcontentsline

\tableofcontents

\vspace{1.cm}

In this Supplemental Material, we provide additional details of the resummation framework for heavy-meson pair photoproduction. In Sec. I, we present the leading-logarithmic one-loop corrections. In Sec. II, we compare resummation results from the unified scheme with those from the conventional massive-only scheme. Section III provides analytic information on the interplay among small-$x$ saturation, the perturbative Sudakov factor, and the NP transverse-momentum broadening, in relation to the heavy-quark mass.

\section{ large logarithms at one-loop diagrams}
This section presents the leading-logarithmic one-loop corrections for heavy-meson–pair photoproduction in the two regimes $|\boldsymbol q| \lesssim m_Q$ and $m_Q \ll |\boldsymbol q|$.  These corrections can be decomposed into an isotropic (angle-independent) term and anisotropic (angle-dependent) harmonic contributions, as follows:
\begin{eqnarray}
\mathcal{S}_{r}(\boldsymbol{k}_g)=
\begin{cases}
\frac{\alpha_{s}}{2\pi^{2}}\frac{1}{\boldsymbol{k}_g^{2}}\left[N_{c}\ln({\boldsymbol{P}^{2}}/{\boldsymbol{k}_g^{2}})+2C_{f}\left(c_{0}(m_Q)+2\sum_{n=1}^{3}c_{2n}(m_Q)\cos\left(2n\phi_{k_gP}\right)\right)\right],
& \boldsymbol{k}_g \leq m_Q,\\[6pt]
\frac{\alpha_{s}}{2\pi^{2}}\frac{1}{\boldsymbol{k}_g^{2}}\left[N_{c}\ln({\boldsymbol{P}^{2}}/{\boldsymbol{k}_g^{2}})+2C_{f}\left(c_{0}(\boldsymbol{k}_g)+2\sum_{n=1}^{3}c_{2n}(\boldsymbol{k}_g)\cos\left(2n\phi_{k_gP}\right)\right)\right],&  \boldsymbol{k}_g \gg m_Q.
\end{cases}
\end{eqnarray}
where $c_{2n}$ (for $n=0,1,2,3$) are given by
\begin{eqnarray}
c_{0} (m_Q)& = & \ln\frac{M_{{\rm in}}^{2}}{m_{Q}^{2}},\\
c_{2} (m_Q)& = & \ln\frac{M_{{\rm in}}^{2}}{m_{Q}^{2}}+f_2(\Delta y_{12})=\ln\frac{M_{{\rm in}}^{2}}{m_{Q}^{2}}+\ln\frac{1}{2(1+\cosh\Delta y_{12})}-1-\frac{1}{2C_{f}N_{c}}\left[1+g_{2}(\Delta y_{12})\right],\\
c_{4} (m_Q)& = & \ln\frac{M_{{\rm in}}^{2}}{m_{Q}^{2}}+f_4(\Delta y_{12})=\ln\frac{M_{{\rm in}}^{2}}{m_{Q}^{2}}+\ln\frac{1}{2(1+\cosh\Delta y_{12})}-\frac{5}{2}-\frac{1}{2C_{f}N_{c}}\left[\frac{1}{2}+g_{4}(\Delta y_{12})\right],\\
c_{6} (m_Q)& = & \ln\frac{M_{{\rm in}}^{2}}{m_{Q}^{2}}+f_6(\Delta y_{12})= \ln\frac{M_{{\rm in}}^{2}}{m_{Q}^{2}}+\ln\frac{1}{2(1+\cosh\Delta y_{12})}-\frac{10}{3}-\frac{1}{2C_{f}N_{c}}\left[\frac{1}{3}+g_{6}(\Delta y_{12})\right],
\end{eqnarray}
with 
\begin{eqnarray}
g_{2}(\Delta y_{12}) & = & \Delta y_{12}\sinh\Delta y_{12}-\cosh\Delta y_{12}\log\left[2(\cosh(\Delta y_{12})+1)\right],\\
g_{4}(\Delta y_{12}) & = & -\Delta y_{12}\sinh(2\Delta y_{12})-2\cosh(\Delta y_{12})+\cosh(2\Delta y_{12})\log\left[2(\cosh(\Delta y_{12})+1)\right],\\
g_{6}(\Delta y_{12}) & = & \Delta y_{12}\sinh(3\Delta y_{12})-\cosh(\Delta y_{12})+2\cosh(2\Delta y_{12})-\cosh(3\Delta y_{12})\log\left[2(\cosh(\Delta y_{12})+1)\right].
\end{eqnarray}
In coordinate space, the harmonic coefficients of the Fourier transformation is still finite due to the additional $\cos(2n\phi_{kgP})$, which can be expressed as
\begin{eqnarray}
  C_{2n}(\mu_{b})= 2n\left(\int_{0}^{m_{Q}}\frac{d\boldsymbol{k}_{g}}{\boldsymbol{k}_{g}}J_{2n}(|b||\boldsymbol{kg}|)\ln\frac{M_{{\rm in}}^{2}}{m_{Q}^{2}} +\int_{m_{Q}}^{\infty}\frac{d\boldsymbol{k}_{g}}{\boldsymbol{k}_{g}}J_{2n}(|b||\boldsymbol{kg}|)\ln\frac{M_{{\rm in}}^{2}}{\boldsymbol{k}_{g}^{2}}+\int_{0}^{\infty}\frac{d\boldsymbol{k}_{g}}{\boldsymbol{k}_{g}}J_{2n}(|b||\boldsymbol{kg}|)f_{2n}(\Delta y_{12})  \right),
\end{eqnarray}
and are given as (for $n=1,2,3$)
\begin{align}
C_{2}(\mu_{b})=&\ln\frac{M_{{\rm in}}^{2}}{\mu_{b,2}^{2}}-\frac{m_{Q}^{2}\boldsymbol{b}^{2}}{8}\,_{2}F_{3}\left(1,1;2,2,3;-\frac{m_{Q}^{2}\boldsymbol{b}^{2}}{4}\right)+f_2(\Delta y_{12}),\\
C_{4}(\mu_{b})=&\ln\frac{M_{{\rm in}}^{2}}{\mu_{b,4}^{2}}-\frac{m_{Q}^{2}\boldsymbol{b}^{2}}{12}\,_{2}F_{3}\left(1,1;2,2,4;-\frac{m_{Q}^{2}\boldsymbol{b}^{2}}{4}\right)-\frac{8J_{2}(|m_{Q}||b|)}{m_{Q}^{2}\boldsymbol{b}^{2}}+1+f_4(\Delta y_{12}),\\
C_{6}(\mu_{b})=&\ln\frac{M_{{\rm in}}^{2}}{\mu_{b,6}^{2}}-\frac{m_{Q}^{2}\boldsymbol{b}^{2}}{16}\,_{2}F_{3}\left(1,1;2,2,5;-\frac{m_{Q}^{2}\boldsymbol{b}^{2}}{4}\right)-6\frac{2J_{2}(|m_{Q}||b|)+4J_{4}(|m_{Q}||b|)}{m_{Q}^{2}\boldsymbol{b}^{2}}+\frac{3}{2}+f_6(\Delta y_{12}).
\end{align}
where $_{2}F_{3}$ is the generalized hypergeometric function and $\ln\mu_{b,2n}^{2}=\ln\mu_{b}^{2}+2\psi^{(0)}(n)+2\gamma_E+1/n$. In the $ \mu_{b,2n}^{2} \lesssim m_Q^2$ region, $C_{2n}(\mu_{b})$ reduces to $\ln(M_{{\rm in}}^{2}/m_{Q}^{2})+f_{2n}(\Delta y_{12})$. By contrast, in the massless limit
only the first (massless) contribution together with the last term suivece, yielding $\ln(M_{{\rm in}}^{2}/\mu_{b,2n}^{2})+f_{2n}(\Delta y_{12})$. These full analytical expressions are well approximated by Eq.~(14) in the main text.

\section{The resummed improved cross-section in the CGC framework}
We apply our unified resummation within the CGC framework, and obtain the resummed improved cross-section as follows
\begin{align}
\frac{d\sigma}{d\Omega} =&
\int \frac{dz_{1h}}{z_{1h}^2} \int \frac{dz_{2h}}{z_{2h}^2} 
\int\frac{d^{2}\boldsymbol{b} }{(2\pi)^{2}}e^{i\boldsymbol{q}\cdot\boldsymbol{b}}e^{-{\rm Sud}(\boldsymbol{b},M_{{\rm in}})} \Big[1+\sum_{n=1} (-1)^n C_{2n}(\mu_b)  \frac{2C_{f}\alpha_{s}(\mu_{b})}{n\pi}\cos(2n\phi_{bP})\Big] D_{h/q}(z_{1h}, \mu_b)D_{h/\bar q}(z_{2h}, \mu_b)
\nonumber \\
 & \times  \alpha_{s}\alpha_{{\rm em}}e_{q}^{2}z(1-z) \int  d^{2}\boldsymbol{q}^\prime e^{-i\boldsymbol{q}^\prime\cdot\boldsymbol{b}} \Big[\mathcal H^{(1)}_{\rm TMD}(z,\boldsymbol{P}){xf}(x_{g},\boldsymbol{q}^\prime) + \cos(2\phi_{Pq^\prime})\mathcal H^{(2)}_{\rm TMD}(z,\boldsymbol{P}){xh}(x_{g},\boldsymbol{q}^\prime)\Big].
\end{align}
with $d\Omega=dy_{1}\,dy_{2}\,d^{2}\boldsymbol{P}_h\,d^{2}\boldsymbol{q}_h$, where $\boldsymbol{p}_{i}$ and $y_i$ denote the transverse momentum and rapidity of each heavy meson, with $\boldsymbol{P}_h=(\boldsymbol{p}_1-\boldsymbol{p}_2)/2$ and $\boldsymbol{q}_h=\boldsymbol{p}_1+\boldsymbol{p}_2$. The transverse momenta of the heavy mesons are defined as $\boldsymbol{p}_{1}=z_{1h}\boldsymbol{k}_{1}$ and $\boldsymbol{p}_{2}=z_{2h}\boldsymbol{k}_{2}$, where $z_{1h}$ ($z_{2h}$) is the fraction of the parent heavy quark’s momentum carried by the $D$ ($\bar D$) meson. At the parton level, $\boldsymbol{k}_i$ denotes the transverse momentum of the heavy quark ($i=1$) and antiquark ($i=2$), with the pair’s total transverse momentum $\boldsymbol{P} = (1-z)\,\boldsymbol{k}_1 - z\,\boldsymbol{k}_2$ and its imbalance $\boldsymbol{q} = \boldsymbol{k}_1 + \boldsymbol{k}_2$, where $z$ is the longitudinal momentum fraction of the heavy quark relative to the incoming photon. Since we focus on forward rapidities and work in the correlation limit $|\boldsymbol P| \gg |\boldsymbol q|$, we take the symmetric configuration $z=1/2$ in the numerical studies. Finally, $x_{\gamma}$ denotes the fraction of the electron (or nucleus) momentum carried by the quasi–real photon, and $x_{g}$ is the small-$x$ gluon momentum fraction in the dense nuclear target. These kinematic variables are explicitly defined as $z=m_{T1}e^{y_1}/(m_{T1}e^{y_1}+m_{T2}e^{y_2})$ and $x_{\gamma/g}=(m_{1T}e^{\pm y_{1}}+m_{2T}e^{\pm y_{2}})/\sqrt{s}$, where the transverse mass is defined as $m_{iT}=\sqrt{\boldsymbol{k}_i^{\,2}+m_Q^2}$, with heavy-quark mass $m_{Q}$ and the collision energy $\sqrt{s}$. In the TMD-factorization framework, for the transverse polarizations of the incoming photon, one obtains an angle-independent term, $\mathcal{H}^{(1)}_{\rm TMD}$, and an azimuthal-modulation term, $\mathcal{H}^{(2)}_{\rm TMD}$, which can be written compactly as follows~\cite{Metz:2011wb}:

\begin{equation}
\mathcal H^{(1)}_{\rm TMD} (z,\boldsymbol{P}) =  \frac{\left(\boldsymbol{P}^{4}+\epsilon_{f}^{4}\right)\left[z^{2}+(1-z)^{2}\right]+2m_Q^{2}\boldsymbol{P}^{2}}{\left(\boldsymbol{P}^{2}+\epsilon_{f}^{2}\right)^{4}}, \quad {\rm and } \quad
\mathcal H^{(2)}_{\rm TMD} (z,\boldsymbol{P})=  \frac{-2\epsilon_{f}^{2}\boldsymbol{P}^{2}\left[z^{2}+(1-z)^{2}\right]+2m_Q^{2}\boldsymbol{P}^{2}}{\left(\boldsymbol{P}^{2}+\epsilon_{f}^{2}\right)^{4}}.
\end{equation}
where $\epsilon_f^{2}=z(1-z)Q^2+m_{Q}^{2}$, with $Q^2$ the photon virtuality. In quasi-real photoproduction, this reduces to $\epsilon_f^{2}=m_{Q}^{2}$. In the Gaussian approximation, the unpolarized $x f$ and linearly polarized $x h$ WW gluon distributions can be written as~\cite{Metz:2011wb}:
\begin{align}
{xf}(x_{g},\boldsymbol{q}^\prime)  =  \mathcal{K} \int d\boldsymbol{r}\boldsymbol{r}\frac{J_{0}(|\boldsymbol{q}^\prime||\boldsymbol{r}|)}{\boldsymbol{r}^{2}}\left[1-\left( S(x_{g},\boldsymbol{r}) \right)^2\right],&\quad {\rm and} \quad  &
{xh}(x_{g},\boldsymbol{q}^\prime)  = \mathcal{K}\int d\boldsymbol{r}\boldsymbol{r}\frac{J_{2}(|\boldsymbol{q}^\prime||\boldsymbol{r}|)}{\boldsymbol{r}^{2}\ln\left(\frac{1}{\boldsymbol{r}\Lambda}+e\right)}\left[1- \left( S(x_{g},\boldsymbol{r}) \right)^2 \right],
\end{align}
with $\mathcal{K}=\frac{S_{\perp}}{2\pi^{3}\alpha_s}\frac{N_{c}^{2}-1}{N_{c}}$ and where $S(x_{g},\boldsymbol{r})$ is the solution of the BK equation. In this work, we obtain it by numerically solving this equation with running coupling corrections (rcBK)~\cite{Golec-Biernat:2001dqn,Balitsky:2006wa, Kovchegov:2006vj, Gardi:2006rp, Balitsky:2007feb,Albacete:2007yr, Albacete:2010sy,Berger:2010sh}.  As initial condition at $x_0=0.01$, we adopt the modified McLerran–Venugopalan model~\cite{Albacete:2010sy} 
\begin{equation}
S(x_{0}=0.01,\boldsymbol{r})=\exp\left[-\frac{\left(Q_{s0}^2 \boldsymbol{r}^2 \right)^{1.118} }{4}\ln\left(\frac{1}{\boldsymbol{r}\Lambda}+e\right)\right].
\end{equation}
We take $\Lambda=0.24~\mathrm{GeV}$ and set the proton’s initial saturation scale to $Q_{s0,p}^2=0.16~\mathrm{GeV}^2$. For a heavy nucleus, we use $Q_{s0,A}^2=5\,Q_{s0,p}^2$~\cite{Shi:2021hwx}.

For the quasi-real photon processes, one should include $x_\gamma f_{\gamma}(x_\gamma)$, which represents the collinear photon flux associated with an electron or a nucleus. For an electron, the photon distribution is
\begin{equation}
f_{\gamma/e}(x_{\gamma})=\frac{\alpha_{\rm em}}{2\pi}\frac{1+(1-x_{\gamma})^{2}}{x_{\gamma}}\ln\!\frac{\mu^{2}}{x_{\gamma}^{2}m_{e}^{2}},
\end{equation}
with the factorization scale $\mu=\mu_b$.
For a nucleus of charge $Z$, the equivalent-photon approximation gives
\begin{equation}
f_{\gamma/A}(x_{\gamma})=\frac{2Z^{2}\alpha_{\rm em}}{\pi}\!\left[\xi K_{0}(\xi)K_{1}(\xi)-\frac{\xi^{2}}{2}\bigl(2K_{1}^{2}(\xi)-K_{0}^{2}(\xi)\bigr)\right],
\end{equation}
where $\xi=2x_{\gamma}m_{p}R_{A}$; $m_p$ stands for the proton mass, $R_A$ denotes the nuclear radius, and $K_{0,1}$ are modified Bessel functions.

\section{Comparison of unified and conventional resummation schemes}

\begin{figure}[!h]
\includegraphics[width=0.44\linewidth]{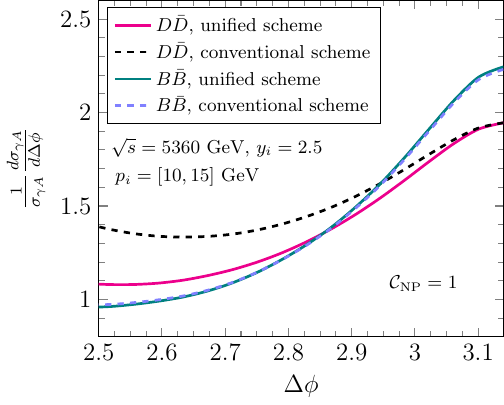}
\hspace{1cm}\includegraphics[width=0.44\linewidth]{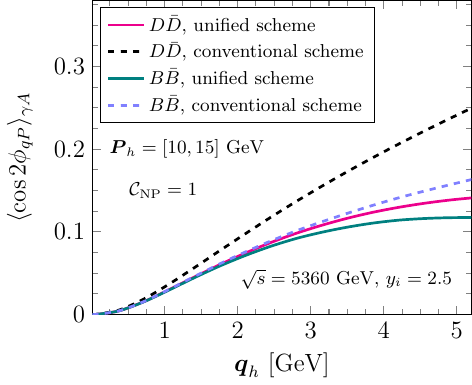}
\caption{Comparison of resummed results from new unified and conventional (massive-only) schemes, for the $\Delta\phi$ correlation (left, as a function of $\Delta\phi$) and the angular asymmetry $\langle \cos(2\phi_{qP})\rangle$ (right, as a function of $\boldsymbol{q}_h$) in D- and B-meson pair photoproduction in UPCs at LHC with $\sqrt{s}=5.36~\mathrm{TeV}$.}
\label{fig:LHC_np025}
\end{figure} 

In this section, as shown in Fig.~\ref{fig:LHC_np025}, we compare resummed predictions with and without matching for D- and B-meson pair photoproduction at $\sqrt{s}=5.36$ TeV. The left (right) panel shows the $\Delta\phi$ correlation (angular asymmetry $\langle \cos 2\phi_{qP}\rangle$). The magenta (teal) solid curves denote the unified resummation results for the D- (B-) meson pairs, whereas the black (blue) dashed curves show the conventional massive-only resummation, valid in the $|\boldsymbol{q}| \lesssim  m_Q$ regime, without matching, for the D- (B-) meson pairs. The left panel of Fig.~\ref{fig:LHC_np025} shows the $\Delta\phi$ distribution for $p_i \in [10,15]$ GeV. For D-meson pairs, the two results coincide in the ultra–back-to-back region; moving away from this limit, they separate, with the unified curve falling more rapidly while the conventional prediction decreases more gradually. For B-meson pairs, the two schemes are indistinguishable over the entire range $\Delta\phi>2.5$. The right panel of Fig.~\ref{fig:LHC_np025} displays $\langle \cos 2\phi_{qP}\rangle$ as a function of $\boldsymbol{q}_h$. For both D- and B-meson pairs, the two results coincide for $|\boldsymbol{q}_h| \lesssim  m_Q/2$, while for larger $\boldsymbol{q}_h$ the unified prediction grows more slowly whereas the conventional prediction increases more rapidly.

These results show that the unified and conventional resummation schemes are consistent for $|\boldsymbol q_h |< m_Q \ll |\boldsymbol P|$, but the unified scheme yields sizable corrections in the region $m_Q \ll |\boldsymbol q_h|$. In addition, the deviation between the two schemes sets in earlier (at smaller ($|\boldsymbol q_h|$)) for higher harmonics ($n$). This demonstrates that the unified scheme is essential for accurate predictions over the full $|\boldsymbol q_h| \ll  |\boldsymbol P|$ domain.

\section{Analytical Analysis}
To gain analytic insight about the interplay among small-$x$ saturation, the perturbative Sudakov factor, and NP broadening, especially in the low-$\boldsymbol q$ region ($|\boldsymbol{q} | <  m_Q$) where saturation effects are maximal, we proceed as follows.

First, starting from the resummed cross section, we define the kernel for the $2n$th azimuthal harmonic, which is proportional to
\begin{eqnarray}
 xF_{2n} (\boldsymbol{q}) \propto  \int\frac{d\boldsymbol{b}}{2\pi} \boldsymbol{b}J_{2n}(|\boldsymbol{q}||\boldsymbol{b}|)x \widetilde f(x_g,\boldsymbol {b}) e^{-{\rm Sud}_i-{\rm Sud}_f},
\end{eqnarray}
where $x\,\widetilde f_{g}(x_g,\boldsymbol b)=x\,f(x_g,\boldsymbol b)\exp(-\mathrm{Sud}_{\mathrm{NP}})$ denotes the modified unpolarized WW gluon distribution, including contributions from both NP broadening and small-$x$ nonlinear evolution. In the limit $|\boldsymbol q|\,|\boldsymbol b|\ll 1$, the Bessel function admits the expansion $J_{2n}(|\boldsymbol q|\,|\boldsymbol b|)\simeq \bigl(|\boldsymbol q|\,|\boldsymbol b|/2\bigr)^{2n}/\Gamma(2n+1)$. Following Refs.~\cite{Shi:2021hwx,Tong:2022zwp,Tong:2023bus}, we then perform a saddle-point approximation of the $\boldsymbol b$-integral, including running-coupling effects, to derive an analytic estimate for the saddle-point scale $\boldsymbol b_{2n}^2$, which is given by
\begin{eqnarray}
\boldsymbol b_{2n}^2=\frac{\Lambda_{\rm QCD}^{2}}{4 \,e^{-2\gamma_E}}
\left[\frac{\boldsymbol{k}^{2} }{\Lambda_{\rm QCD}^{2}}\left(\frac{M_{{\rm in}}^{2}}{m_{Q}^{2}}\right)^{\frac{2C_{f}}{N_{c}}} \right]^{\frac{-1}{\left(n+1\right)\beta_{0}+1}}.
\label{eq::b2n}
\end{eqnarray}
We observe that this scale $\boldsymbol b_{2n}^2$ is governed by the perturbative Sudakov factor, from the initial and final emissions, and by the harmonic order $n$. Furthermore, our saddle-point scale closely matches that reported in Ref.~\cite{Shi:2021hwx} for the $n=0$ case and that found in Ref.~\cite{Tong:2022zwp} in $n \geq 0$.

Second, we examine how the harmonics vary with the total broadening scale $(Q_s^2+Q_{\rm NP}^2)$, harmonic order $n$, and quark mass $m_Q$. After employing the saddle-point approximation, the harmonics are proportional to the ratio of $x \widetilde f_{g}(x_g,\boldsymbol b)$ evaluated at different $\boldsymbol b$. In the $\boldsymbol b^2\ll 1/(Q_s^2+Q_{\rm NP}^2)$ limit, the Fourier transform is dominated by the large-$\boldsymbol q$ tail of the WW distribution, $\tilde x f(x_g,\boldsymbol q)$, in the $\boldsymbol q^2\gg (Q_s^2+Q_{\rm NP}^2)$ regime, which yields $x \widetilde f(x_g, \boldsymbol b)\propto Q_s^2 \ln \!\bigl[{1}/{(Q_s^2+Q_{\rm NP}^2)\boldsymbol b^2}\bigr]$, where $Q_{\rm NP}$ parametrizes combined NP-physics effects, and we assume the same value of $Q_{\rm NP}^2$ for both proton and nuclear targets. After applying the saddle-point value in the cross section, the angular asymmetries $\langle \cos 2n\phi_{qP}\rangle$ behave as follows
\begin{equation}
\langle \cos2n\phi _{qP}\rangle  \propto c_{2n}
\frac{x \widetilde f(x_g,\boldsymbol{b}_{2n})}{x  \widetilde f(x_g, \boldsymbol{b}_0)} \simeq  c_{2n}  \frac{\ln [{(Q_s^2+Q_{\rm NP}^2) \boldsymbol{b}_{2n}^2}]}{\ln [{(Q_s^2+Q_{\rm NP}^2) \boldsymbol{b}_0^2}]}.
\end{equation} 
Differentiating the azimuthal asymmetry with respect to the total broadening scale $(Q_s^2+a_{\rm NP}^2)$ gives
\begin{equation}
\frac{\partial \langle \cos2n\phi _{qP}\rangle }{ \partial (Q_s^2+Q_{\rm NP}^2) } 
\propto 
 \ln \frac{M_{\rm in}^2}{m_Q^2} 
 \frac{\ln \left( \boldsymbol{b}_{0}^2/\boldsymbol{b}_{2n}^2 \right) }{\ln ^2 \left[  (Q_s^2+Q_{\rm NP}^2) \boldsymbol{b}_0^2\right]}<0,
\end{equation}
so that $\langle \cos2n\phi _{qP}\rangle$ is a strictly decreasing function of the total broadening scale $(Q_s^2+Q_{\rm NP}^2)$. Furthermore, $\langle \cos2n\phi _{qP}\rangle$ falls off monotonically with increasing $\boldsymbol b$ and carries the positive prefactor $\ln(M_{\rm in}^2/m_Q^2)$. Because $\boldsymbol b_{2n}^2>\boldsymbol b_{2(n-1)}^2>\boldsymbol b_{0}^2$ for $n\geq 2$, the harmonics are further suppressed for higher harmonic order $n$, for larger saturation scale $Q_s^2$ or larger NP scale $Q_{\rm NP }^2$, and for heavier quark mass $m_Q$.

Next, we examine how the harmonic ratio for a nuclear target to that for a proton target varies under the competing broadening effects: saturation, the NP broadening, and the perturbative Sudakov effect. This harmonic ratio is proportional to $x \widetilde f_{A}/x \widetilde f_{p}$ evaluated at different $\boldsymbol{b}$. $x \widetilde f_{A}/x \widetilde f_{p}$ can be written as follows
\begin{equation}
\frac{x \widetilde f_A(x_g,\boldsymbol{b})}{A^{1/3} x \widetilde f_p(x_g, \boldsymbol{b})}  \simeq 
\frac{\ln[\left( A^{1/3} Q_{sp}^2 +Q_{\rm NP}^2  \right) \boldsymbol{b}^2]}{\ln [\left( Q_{sp}^2 +Q_{\rm NP}^2  \right) \boldsymbol{b}^2]},
\end{equation}
where we simply assume that $Q_{sA}^2= A^{1/3} Q_{sp}^2$. In the regime where NP broadening far exceeds the saturation scale, the modified unpolarized WW gluon reduces to $x \widetilde f(x_g, \boldsymbol{b}) \big|_{Q_{\rm NP}^2 \gg Q_s^2}  \propto Q_{\rm s}^2\ln [{1  }/({Q_{\rm NP}^2 \boldsymbol{b}^2})]$ and the TMD become effectively target-independent, so that $x \widetilde f_A/x \widetilde f_p \simeq  A^{1/3}$ and the NP physics dilutes the signature of the saturation effect.

 By contrast, when the saturation scale dominates, one finds $ x \widetilde f(x_g, \boldsymbol{b}) \big|_{Q_{\rm NP}^2 \ll Q_s^2}  \propto Q_s^2\ln [{1  }/({Q_s^2 \boldsymbol{b}^2})]$. Differentiating the nuclear‐to‐proton WW gluon ratio $x\widetilde f_A/x\widetilde f_p $ with respect to either $\boldsymbol{b}^2$ or $Q_{sp}^2$ gives
\begin{eqnarray}
\frac{\boldsymbol{b}^2  \partial \frac{x \widetilde f_A(x_g,\boldsymbol{b})}{x \widetilde f_p(x_g, \boldsymbol{b})} }{ A^{1/3}  \partial \boldsymbol{b}^2 }=
\frac{Q_{sp}^2\partial \frac{x \widetilde f_A(x_g,\boldsymbol{b})}{x \widetilde f_p(x_g, \boldsymbol{b})} }{A^{1/3}  \partial Q_{sp}^2 } = \frac{-\ln A^{1/3}}{\ln^2 (Q_{sp}^2 \boldsymbol{b}^2)}<0,
\label{eq::Rb2n}
\end{eqnarray}
showing that the ratio $x\widetilde f_A/x\widetilde f_p $ decreases as the transverse size $\boldsymbol{b}^2$ or the saturation scale $Q_{sp}^2$ increases. These results imply that the nuclear suppression becomes increasingly stronger at higher harmonic order $n$, as larger $n$ weights the distribution toward larger $\boldsymbol{b} $. Conversely, increasing the strength of the perturbative Sudakov factor shifts the dominant $\boldsymbol{b}$ region to smaller values, thereby leads to less visible saturation effects.

Finally, we discuss the mass hierarchy. From Eq.~(\ref{eq::b2n}), a heavier quark mass $m_Q$ leads to a smaller $\boldsymbol b_{2n}$ (i.e., $\boldsymbol b_{2n,c}^2>\boldsymbol b_{2n,b}^2$), implying a weaker perturbative Sudakov effect, consistent with dead-cone suppression of soft radiation. Consequently, the Sudakov-induced $p_\perp$ broadening is weaker for B-meson pairs than for D-meson pairs; correspondingly, the $\Delta\phi$ distribution for $B\bar B$ is more sharply peaked (left panel of Fig.~\ref{fig:LHC_np025}). Moreover, according to Eq.~(\ref{eq::Rb2n}), a smaller $\boldsymbol b_{2n}$ yields a smaller nuclear-to-proton WW gluon ratio, such that
$x\widetilde f_A/x\widetilde f_p |_{b=\boldsymbol b_{2n,c}}>x\widetilde f_A/x\widetilde f_p |_{b=\boldsymbol b_{2n,b}}$
Thus, heavier quark masses enhance the visibility of saturation signatures. As shown in Fig.~\ref{fig:LHC_dphi_R} of the main text, the nuclear modification factor $\mathcal R$ in the $\Delta\phi$ distribution exhibits a clear mass ordering,
$\mathcal{R}\big|_{m_c} > \mathcal{R}\big|_{m_b}$.

For the azimuthal harmonics, $\langle \cos (2n \phi_{qP})\rangle$ ratios between $\gamma A$ and $\gamma p$ collisions (see the lower panel of Fig.~\ref{fig:D_cos2nphi_UPC}) retain a clear mass ordering, since heavier quarks further suppress soft-gluon radiation. We therefore conclude that increasing the quark mass enhances the visibility of saturation effects.

Ultimately, we have elucidated the interplay among true small-$x$ saturation dynamics, NP transverse-momentum broadening, and perturbative soft-gluon radiation. While both NP broadening and the perturbative Sudakov factor tend to dilute saturation signatures in the lower harmonics, the higher-order angular coefficients remain primarily shaped by nuclear suppression due to gluon saturation. In addition, heavier quark masses make saturation signatures more pronounced, since heavy-quark mass effects suppress soft-gluon radiation. Consequently, measurements of large-$n$ azimuthal harmonics offer a robust window into genuine nonlinear gluon dynamics at future EIC and UPC experiments.

\end{widetext}

\end{document}